\documentclass[a4paper,twocolumn,english,american,prl,superscriptaddress,showpacs]{revtex4}
\usepackage[T1]{fontenc}
\usepackage[latin1]{inputenc}
\usepackage{graphicx}
\usepackage{amssymb}

\makeatletter

\makeatletter
\usepackage{ae,aecompl}
\usepackage{dcolumn}
\usepackage[normalem]{ulem}
\usepackage{psfrag}

\def\kbar{{\mathchar'26\mkern-9mu k}}

\usepackage{babel}
\makeatother

\usepackage{babel}
\makeatother
\begin{document}

\title{Quantum scaling laws in the onset of dynamical delocalization}

\author{Julien Chabé}

\author{Hans Lignier}
\altaffiliation[Present address: ]{Dipartimento di Fisica ``E. Fermi", Universit{\`a} di Pisa, Pisa, Italy}

\author{Hugo L. D. de Souza Cavalcante}
\altaffiliation[Present address: ]{Departamento de F{\'i}sica, Universidade Federal de Pernambuco, Recife, Brazil}

\affiliation{Laboratoire de Physique des Lasers, Atomes et Molécules, UMR CNRS
8523, Université des Sciences et Technologies de Lille, 
Centre d'Etudes et Recherches Lasers et Applications, 
F-59655 Villeneuve d'Ascq Cedex,
France}

\homepage{www.phlam.univ-lille1.fr}

\author{Dominique Delande}

\affiliation{Laboratoire Kastler-Brossel, Case 74, UPMC, 4 Place Jussieu, F-75252 Paris Cedex 05, France}

\author{Pascal Szriftgiser}

\author{Jean Claude Garreau}

\affiliation{Laboratoire de Physique des Lasers, Atomes et Molécules, UMR CNRS
8523, Université des Sciences et Technologies de Lille, 
Centre d'Etudes et Recherches Lasers et Applications, 
F-59655 Villeneuve d'Ascq Cedex,
France}

\date{\today}

\begin{abstract}
We study the destruction of dynamical localization,
experimentally observed in an atomic realization of the kicked rotor,
by a deterministic Hamiltonian perturbation, with a temporal
periodicity incommensurate with the principal driving. We show 
that the destruction is gradual,
with well defined scaling laws for the various classical and quantum parameters,
in sharp contrast with predictions based on the analogy with Anderson localization.
\end{abstract}


\maketitle

Quantum chaos is defined as the dynamical behavior of a \emph{quantum}
system whose \emph{classical} limit is chaotic. This has triggered
a large number of studies trying to relate classical properties
to quantum properties, e.g. Lyapunov exponents to quantum fidelity
\cite{Jalabert:Loschmidt:PRL01,Zurek:Loschmidt:PRL03}, or to detect
quantum stability in a quantum-chaotic system \cite{DArcy:QuantumStability:PRL03}.

Quantum-chaotic dynamics manifest themselves by characteristic behaviors
in which quantum interference plays an important role, making the
dynamics distinct from classical dynamics. An example,
that shall concern us particularly here, is \emph{dynamical localization}
(DL) \cite{Casati:LocDyn:79,Izrailev:LocDyn:PREP90}, observed in
time-periodic systems. DL is the suppression of the classical chaotic diffusion
by quantum interference due to long-range coherence in momentum space;
it manifests itself after a typical ``localization time'' as an
exponential localization of the average momentum distribution. 
Because the system
is time-periodic, one can use the Floquet theorem to build a basis
of quasi-eigenstates (states that are left unchanged, except for a phase factor, 
under the temporal evolution over one period). This makes it
possible to map the quasi-eigenstates
of the time-periodic system on the true eigenstates of a quasi-random
static one-dimensional system, which presents the non-trivial Anderson
localization. 
Anderson (or strong) localization has been a major subject in physics
in the last decades, with implications in several areas, beyond the
primary field of solid state physics
~\cite{Anderson:LocAnderson:RMP78,Fishman:LocDynAnderson:PRA84,Abrahams:Scaling:PRL79,Kramer:Localization:RPP93}. 
In this paper, we show
that studying the breakdown of dynamical localization may also
bring some new insight on the physics of Anderson localization.
The latter is known to be strongly dependent on the
number of freedoms, with marginal localization in dimension
2 and the coexistence of localized and delocalized states -- depending
on the parameters -- in dimension 3. By playing with the temporal
dependance of the Hamiltonian -- for example by adding incommensurate
frequencies to make a quasi-periodic Hamitonian -- it is possible
to study temporal equivalents of the Anderson model in various dimensions.
What happens if we introduce progressively a second (incommensurate) frequency in
the system, by increasing its strength from zero? As the system
is quasi-periodic with two basic frequencies, it is reasonnable to
assume that it can be mapped onto a two-dimensional Anderson model~\cite{Shepel:Bicolor:PD83,DoronFish:AndLoc:PRL88},
which, for a small perturbation, is a quasi-1D model, and one could then expect
localization to be preserved, at the cost of an increased
localization length. Theoretical studies based on the analogy with Anderson localization,
supplemented by numerical simulations, indeed predict that the onset
of dynamical delocalization takes place when a quasi-periodic perturbation
with finite \emph{non-zero} strength is applied on the 
system~\cite{Casati:IncommFreqsQKR:PRL89}. In the present work, 
we show experimentally that
this is \emph{NOT} the case, and that DL is destroyed as soon as 
the pertubation is non zero, and unravel the scaling laws which 
govern the phenomenon.

We consider an atomic version of the kicked rotor,
a paradigmatic system for theoretical and experimental studies of
classical \cite{Chirikov:ChaosClassKR:PhysRep79} and quantum chaos
\cite{Raizen:LDynNoise:PRL98,Christ:LDynNoise:PRL98,DArcy:AccModes:PRL99,AP:Bicolor:PRL00},
which consists in exposing laser-cooled
atoms to short, periodic pulses of a far-detuned standing wave, so
as to obtain an atomic equivalent of the kicked rotor. Using this
system, DL has been unambiguously observed and its characteristics
studied \cite{Raizen:LDynFirst:PRL94}.
The temporal periodicity is a key ingredient. For
example, random fluctuations on the strengths of the successive kicks, 
have been experimentally shown to destroy DL~\cite{Raizen:LDynNoise:PRL98},
even for fluctuations not significantly affecting the
classical diffusive behaviour. Similarly, the introduction of a small
amount of non-Hamiltonian evolution -- spontaneous emission and the
associated random recoil of the atom in the experiment 
\cite{Raizen:LDynNoise:PRL98,Christ:LDynNoise:PRL98}
-- is enough to induce decoherence, and thus to reduce or kill quantum
interference effects, thus restoring the classical dynamics. 

In previous works, we have extended the study of the kicked rotor to the
two-frequency quasiperiodic case by adding a second series of kicks: the laser-cooled
atoms interact with a modulated standing wave of wavevector 
$\mathbf{k}_{L}=k_{L}\mathbf{x}$
forming two series of kicks at frequencies $f_{1}=1/T_{1}$ (primary
series) and $f_{2}=rf_{1}$ (secondary series), so as to obtain a
system described by the Hamiltonian:
\begin{equation}
H=\frac{P^{2}}{2}+\sin\theta\left[K\sum_{n=0}^{N-1}\delta(t-n)+
aK\sum_{n=0}^{rN-1}\delta\left(t-\frac{n}{r}\right)\right]
\label{eq:Hbicolor}
\end{equation}
 where we measure momentum in units of $2\hbar k_{L}$, $\theta=2k_{L}x$,
time in units of $T_{1}$. The normalized kick amplitude is 
$K=\Omega_{1}^{2}\hbar k_{L}^{2}\tau T_{1}/(2M\Delta)$
($\Omega_{1}$ is the resonant Rabi frequeny, proportional to the
light intensity and $\tau$ is the duration of the kicks 
\footnote{In the real experiment the standing wave pulses have a finite
duration of $\tau=600$ ns (of square shape). They can be considered
as $\delta$ functions as far as the motion of the atom during the
pulse is negligible compared to the spatial period of the potential,
that is for $\left\langle p\right\rangle \tau/M\ll\lambda_{L}/2$
($p$ is the momentum and $M$ is the mass of the atom), condition
that is easily satisfied in the experiment.}. 
In such units, the normalized Planck constant, describing the
``quanticity'' of the system, is $\kbar=4\hbar k_{L}^{2}T_{1}/M$;
it can thus be controlled by changing the frequency of the kicks.
We have shown that, in the quasiperiodic case ($r$ irrational), with
$a=1,$ DL is destroyed \cite{AP:Bicolor:PRL00}.

What are the scaling laws for the onset of delocalization? In order to understand 
the origin of such laws, we analyze pertubatively the effect
of the second series. 
The effect of each individual kick is 
expressed by a unitary evolution operator:
\begin{equation}
U(a,K,\kbar) = \exp{\left( - i \frac{a K \sin \theta}{\kbar} \right) }.
\end{equation}

For sufficiently small $a$ -- such that $aK/\kbar\ll 1$ -- this operator is close
to unity and a single kick only slightly modifies the atomic state.
It is the accumulation of a series of small kicks which significantly 
perturbs the dynamics. If the ratio $r$ of the two frequencies is sufficiently
far from any simple rational number, the second series of kicks
is applied at quasi-random phases (measured with respect to the principal
sequence), so that there is no coherent action of consecutive kicks.
In classical language, the positions $\theta$
at consecutive secondary kicks are uncorrelated. 
In the unperturbed Floquet basis, the \emph{incoherent} cumulative 
effect of the secondary kicks results in a diffusive process.
The strength of a single kick being proportional to 
$aK/\kbar,$ the incoherent cumulative effect of $n$ secondary kicks is
proportional to $na^2K^2/\kbar^2,$ and the characteristic
time scale for the effect of the secondary kick series then scales as
$T_2 \kbar^2/(a^2K^2).$ The other important time scale in the problem
is the localization time, scaling like $T_1 K^2/\kbar^2.$ 
If over one localization time,
the effect of the second kick sequence is small, DL has time
to establish before being destroyed. In the opposite situation, diffusion in the Floquet
basis is the dominant process and no localization is expected.
The cross-over between the two regimes arrives when the two time scales are comparable,
i.e. when $T_2 \kbar^2/(a^2K^2) \simeq T_1 K^2/\kbar^2$, or
(assuming $r$ is of the order of unity):
\begin{equation}
\tilde{a}=\frac{aK^{2}}{\kbar^{2}}\simeq 1
\label{tildea}
\end{equation}
$\tilde{a}$ thus represents the scaled parameter 
governing the onset of delocalization. It depends on both the
``chaoticity" parameter $K$ and the ``quanticity" parameter, 
the effective Planck constant $\kbar,$
which shows the intrinsic quantum nature of the phenomenon.
Note that the preceding discussion establishes the expression
for the relevant parameter
$\tilde{a},$ but is not sufficient for knowing whether there
is an abrupt (as predicted from the Anderson model) 
or a smooth transition (as we experimentally observe here) between
localization and delocalization. 

The experimental setup has been described in detail elsewhere 
\cite{AP:ChaosQTransp:CNSNS:2003,AP:RamanSpectro:PRA02,AP:DiodeMod:EPJD99}. 
Cesium atoms are
first trapped and cooled in a standard Magneto-Optical trap, down
to a temperature around 3 $\mu$K. The trap is turned off, and the
atoms interact with the doubly-pulsed standing wave [Eq.~(\ref{eq:Hbicolor})]. 
Raman stimulated
transitions are then used to measure the population $\Pi(P)$ of a
given momentum class, which can be chosen by changing the Raman detuning.
It is very easy to directly measure the degree of localization of
the system by measuring the population in the zero momentum class $\Pi_{0}=\Pi(P=0)$.
As the number of atoms in a given experiment is constant,
this value is smaller if the momentum distribution is larger, that
is $\Pi_{0}\propto\left\langle \Delta P^{2}\right\rangle ^{-1/2}.$
In practice, in order to improve the signal to noise ratio, we measure
the fraction of atoms with velocity in a small range around zero.
The range is comparable to the width of the initial velocity distribution
(few recoil velocities) and much smaller than the final width. In
our experiment, the standing wave is obtained from a SDL MOPA (Master
Oscillator Power Amplifier) delivering around 350 mW. The beam is transported
through polarization-maintaining optical fibers to the interaction
region. A diode laser mounted in extended cavity configuration and
locked on an invar Fabry-Perot interferometer serves as master. The
frequency is continuously
monitored by an Advantest Q8326 lambda-meter. 

\begin{figure}
\begin{centering}
\includegraphics[clip,width=8cm]{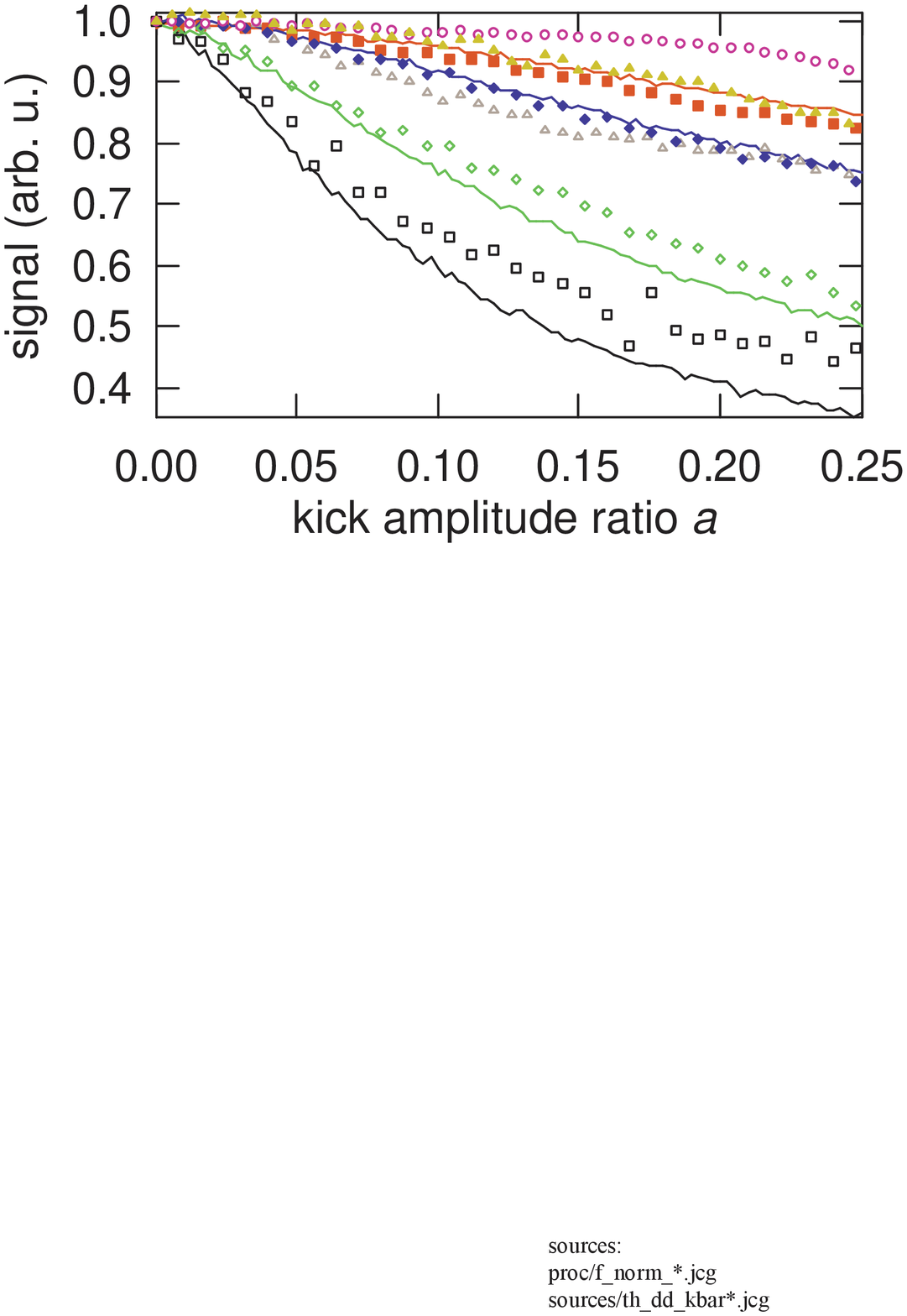}
\end{centering}
\caption{\label{fig:ExpLocDegrees} 
(Color online) Normalized number of zero-velocity atoms
as a function of the amplitude ratio $a$ of the secondary sequence
of kicks to the principal sequence of kicks for various values of
$\kbar$ and $K$ . In order to ease the comparison, the curves have
been normalized so that the value at $a=0$ is 1 for all curves. 
The parameters and plotting conventions are listed in 
table~\ref{tab:ParameterSets}.
The solid lines
are numerical simulations for the four curves at $K=6.8,$
with no adjustable parameter, as all quantities
of interest have been measured experimentally. The kick strength
at the center of the atomic cloud is calculated from the laser power
and the geometry of the beam, the temporal profile of the kicks is
measured and spontaneous emission rates are calculated from the laser
intensity, the detuning and the oscillator strength of the atomic
transition. }
\end{figure}

In order to study the destruction of DL, we choose a number of
kicks that is larger than the localization time $N_{L}\propto(K/\kbar)^{2}$
and measure $\Pi_{0}$ for increasing values of $a$ from zero to
0.25. Fig.~\ref{fig:ExpLocDegrees} displays the typical results for
seven sets of parameters, that are shown in table~\ref{tab:ParameterSets}.

Numerical simulations of the kicked rotor quantum dynamics are useful for
a detailed interpretation of the results.
A few complications must be included in our simulations, which are, ordered
by decreasing importance: the finite temporal length of the pulses
which makes the kicks slightly different from $\delta$-kicks, the
spatial variation of the laser intensity across the atomic cloud which
implies that all the atoms do not feel the same $K$ value 
and some residual spontaneous emission. Altogether, they affect
the shape of the curves in a rather limited way: the decay
of $\Pi_0$ with $a$ is slower by about 20\%. 
Fig.~\ref{fig:ExpLocDegrees} shows a comparison of the numerical calculation 
for the four curves at $K=6.8$ and various $\kbar$ values
with the corresponding experimetal curves: 
the agreement is very
good. There is no adjustable parameter, all the quantities have been
either directly measured or calculated from measured quantities and
first principles. Small deviations are observed for the lowest curve.
This is probably an experimental artefact due to the long duration
of the full kick sequence (2.8 ms, compared to 1.2 ms for the upper curve).
The atoms are freely falling due to gravity, and, as they escape the
central region of the Gaussian-profiled standing wave, they see a
smaller light intensity and the momentum diffusion is stopped, and
thus the delocalization. For the lowest curve, the motion of the atoms
is almost 6 times larger than for the highest curve. 

In the various series, a constant ratio $N/N_{L}=2.5$ ($N_L$ is the
localization time) is kept in
order to insure that all experiments correspond to the same ``localization
stage''. The various sets of parameters have been chosen to allow
us to vary either $\kbar$ or $K$ keeping all other parameters constant. 
We choose $r=0.681$,
a typical ``irrational" number, i.e. far from any simple
rational, in order to avoid DL and sub-Fourier 
resonances~\cite{AP:SubFourier:PRL02} which are observed in a 
narrow range around rational numbers. 

\begin{figure}
\begin{centering}
\includegraphics[clip,width=8cm]{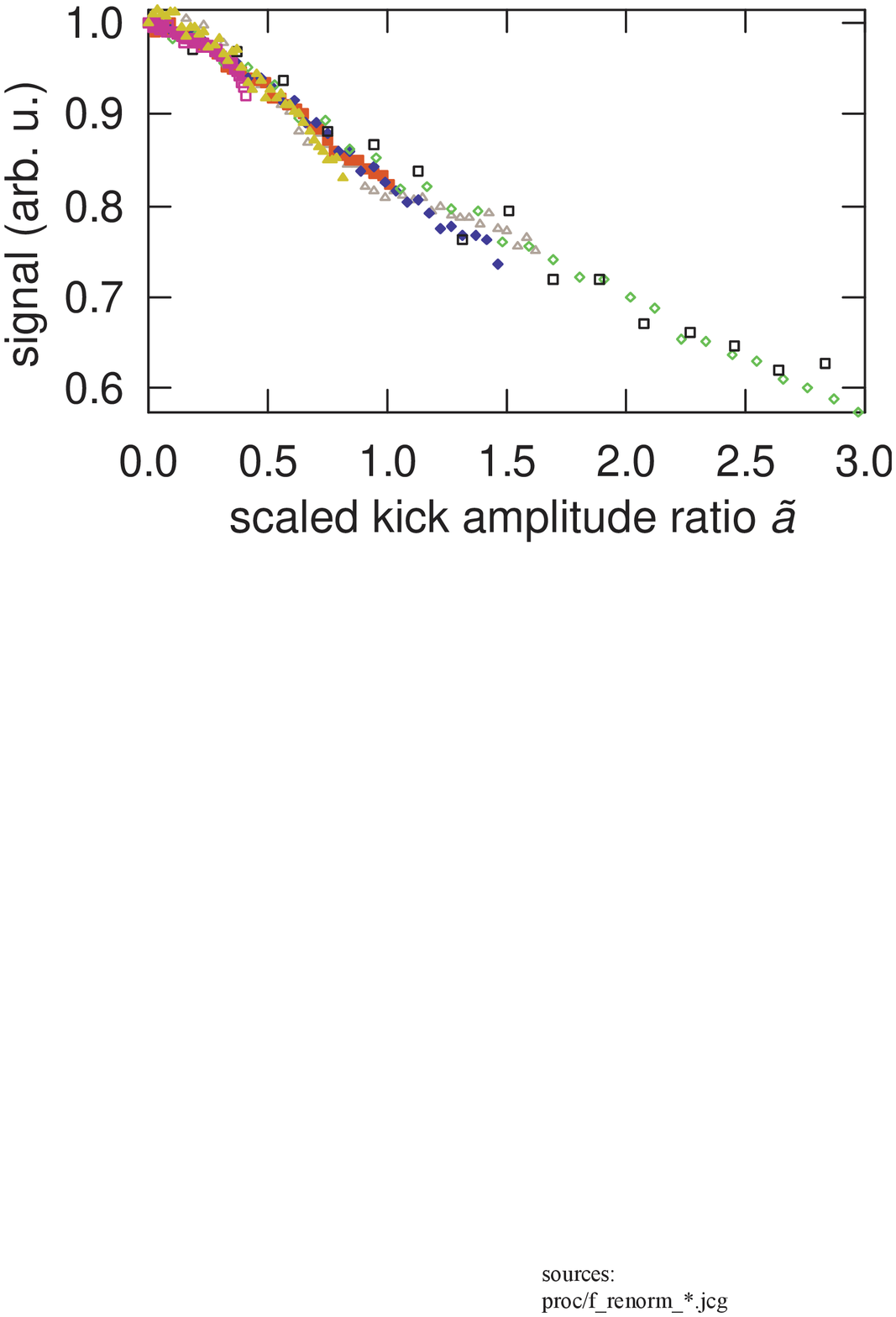}
\end{centering}
\caption{\label{fig:rescaled}
(Color online) Number of zero-velocity atoms as a function
of the scaled kick series amplitude ratio, $\tilde{a}=aK^{2}/\kbar^{2}$.
One observes the superposition of all curves. Parameters and graphic
conventions are the same as in Fig.~\ref{fig:ExpLocDegrees}.}
\end{figure}

Data in Fig.~\ref{fig:ExpLocDegrees} clearly demonstrate that the
destruction of DL by a second series of kicks is gradual and certainly
not a phase transition. The various curves display a qualitatively
similar behaviour, a signature of universality in the destruction
of DL. In order to exhibit this universal behaviour, we show in 
Fig.~\ref{fig:rescaled} the same data plotted as a function of the 
scaled amplitude $\tilde{a},$ given by Eq.~(\ref{tildea}). 
The seven experimental curves now coincide, which proves that $\tilde{a}$
is the truly relevant parameter. 

\begin{table}
\begin{ruledtabular}
\begin{tabular}{cccccccc}
$f_1$ (kHz) & $\Delta$ (GHz) & $P$ (mW) & $N$ & $N_L$ & $K$ & $\kbar$ & Symbol \\
\hline
30.000  & -18.8 & 95 & 35 & 14 & 6.8 & 3.46 & $\blacksquare$ \\
36.000  &  -15.6 & 95 & 50 & 20 & 6.8 & 2.88 & $\blacklozenge$ \\
54.000  & -10.5 &  95 &113 & 45 & 6.8 & 1.92 & $\lozenge$\\
72.000  & -7.9 & 95 & 200 & 79 & 6.8 & 1.44 & $\square$\\
30.000   & -21.3 & 62 &18 & 7 & 4.5 & 3.46 & $\bigcirc$\\
30.000   & -21.3 & 87 & 35 & 14 & 6.3 & 3.46 & $\blacktriangle$\\
30.000   & -21.3 &123 & 70 & 28 & 8.9 & 3.46 & $\triangle$\\
\end{tabular}
\end{ruledtabular}
\caption{\label{tab:ParameterSets}
Sets of parameters used in the curves of
Fig.~\ref{fig:ExpLocDegrees}. The parameters $r=f_{2}/f_{1}=0.681$
and the ratio $N/N_{L}\approx2.5$ ($N_L$ is the localization
time) are the same for all data series.
The pulse duration $\tau$ is 0.6 $\mu$s for the 4 top lines and
0.7 $\mu$s for the bottom 3 ones.}
\end{table}

\begin{figure}
\begin{centering}
\includegraphics[clip,width=8cm]{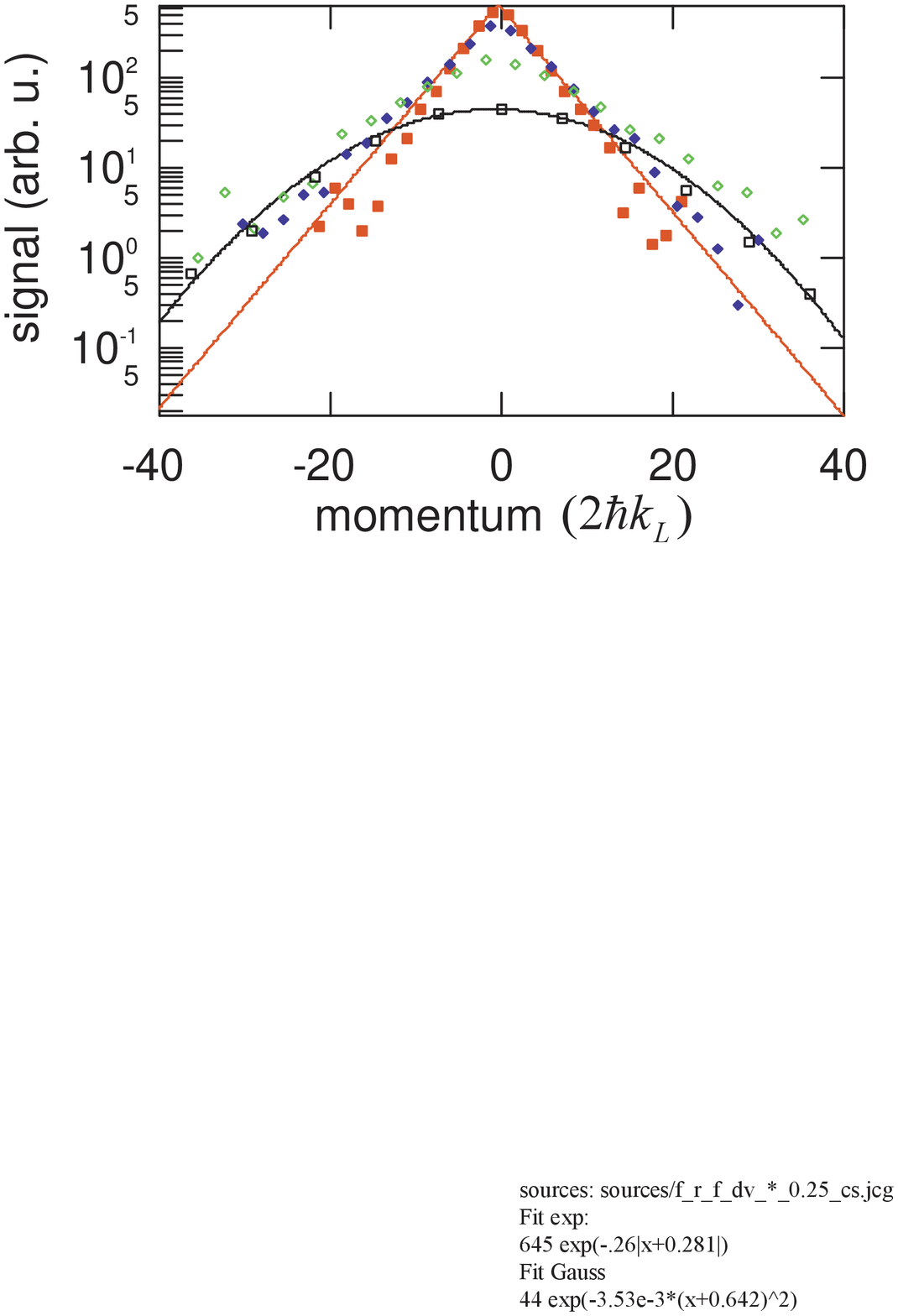}
\end{centering}
\caption{\label{fig:MomentumDistributions}
(Color online) Experimentally observed momentum distributions at $a=0.25$
in log scale, for various values of $\kbar$ and constant $K=6.8$.
The curve for $\kbar=3.46$ which corresponds to $\widetilde{a}=0.97$
is well fitted by an exponential, whereas the curve with $\kbar=1.44$,
or $\widetilde{a}=5.6$ is fitted by a Gaussian.}
\end{figure}

Although the preceeding results are clear-cut proofs that the second
series of kicks gradually reduce the localization of the system, this 
may result from two completely different mechanisms: either the second
series destroys the DL and restores a diffusive behavior of the quantum
system, with a diffusion constant smoothly increasing from zero (for
vanishing $a$), or the localization is preserved, but with a localization
length smoothly increasing with $a.$ Which of two scenarios, the
diffusive scenario or the Anderson scenario, correctly describes the
physics at work, cannot be decided from the preceding results. This
issue can be solved by looking at the momentum distribution. Indeed,
the Anderson localized regime is characterized by an exponential localization
of the wavefunction in momentum space, while the diffusive regime
is associated with a Gaussian momentum distribution. 
Fig.~\ref{fig:MomentumDistributions}
shows the experimentally observed momentum distribution for various
values of $\kbar$ and $K=6.8$. DL is clearly observed for the smaller
value of the scaled amplitude $\widetilde{a}=0.97$ ($\kbar=3.46$,
exponential shape) whereas the larger value $\widetilde{a}=5.6$ is
clearly in the diffusive regime ($\kbar=1.44$, Gaussian
shape). The two other plots present intermediate shapes between a Gaussian
and an exponential. We thus conclude that the diffusive scenario is
the correct one. This is somehow surprising, as theoretical arguments
and numerical calculations~\cite{Casati:IncommFreqsQKR:PRL89,Shepel:Bicolor:PD83,DoronFish:AndLoc:PRL88} 
on a slightly different
quasi-periodic system -- where a single series of kicks has a periodically
modulated (at a incommensurate frequency) amplitude -- show that the
Anderson scenario applies. A possible explanation of this apparent
incompatibility might be that quasi-periodicity with two incommensurate
frequencies in the driving of the system is not sufficient to determine
whether the system is localized or not. In other words, quasi-periodic
driving of a Hamiltonian system might not lead to universal behaviour.
This is a rather difficult theoretical problem, never treated in the
litterature, to the best of our knowledge. Experiments on the quasi-periodically 
driven atomic rotor may help to clarify this stimulating issue.

In conclusion, we have observed that the destruction of dynamical
localization by the addition of a small Hamiltonian periodic perturbation
at a frequency incommensurate with the principal driving, leads to
a gradual destruction of the localization, through a continuous growth
of a residual diffusion constant, and \emph{NOT} to the equivalent
of Anderson localization in a two degrees of freedom system. We have
also determined and tested the quantum scaling laws governing the onset of delocalization.

\begin{acknowledgments}
Laboratoire de Physique des Lasers, Atomes et Molécules (PhLAM) is
Unité Mixte de Recherche UMR 8523 du CNRS et de l'Université des Sciences
et Technologies de Lille. Laboratoire Kastler Brossel is laboratoire
de l'Université Pierre et Marie Curie et de l'Ecole Normale Supérieure,
UMR 8552 du CNRS. CPU time on various computers has been provided
by IDRIS.
\end{acknowledgments}

\end{document}